\documentclass[aps,preprintnumbers,eqsecnum,amsmath,amssymb,nofootinbib]{revtex4}  %eqsecnum,
\usepackage{graphicx} 
\usepackage{bm} 

\begin{document}
%\normalsize

\title{Three-particle distribution in {\it B} meson and charm-quark loops in FCNC {\it B} decays}
\author{Dmitri Melikhov}
\affiliation{ 
D.~V.~Skobeltsyn Institute of Nuclear Physics, M.~V.~Lomonosov Moscow State University, 119991, Moscow, Russia;\\
Joint Institute for Nuclear Research, 141980 Dubna, Russia;\\
and Faculty of Physics, University of Vienna, Boltzmanngasse 5, A-1090 Vienna, Austria
}
\begin{abstract}
We discuss a nonfactorizable (NF) contribution of a charm loop to the FCNC $B$-decay amplitude
given through the three-particle Bethe-Salpeter amplitude (3BS) of the $B$-meson.
This 3BS contains one heavy-quark field and two light fields (a light quark and a gluon). 
Our discussion is aimed at clarifying properties of the $B$-meson 3BS 
necessary to describe properly charm-loop contributions to the amplitudes of FCNC $B$-decays.
We demonstrate that the dominant contribution of nonfactorizable charm to FCNC $B$-decay amplitude
is given in the heavy-quark limit by a convolution of some hard kernel and the $B$-meson 3BS in a ``double-collinear''
light cone (LC) configuration: one of the light degrees of freedom $\phi(x)$, $x^2=0$, lies on the
$(+)$-direction of the LC, whereas another light degree of freedom $\phi'(x')$, $x'^2=0$ lies along
the $(-)$-direction. We show the emergence of new constraints on the distribution amplitudes which parametrize
the 3BS in this double-collinear configuration. 
\end{abstract}
\date{\today}
\maketitle
\normalsize
\section{Introduction}
\label{Sec_introduction}
Charming loops in rare flavour-changing neutral current (FCNC) decays of the $B$-meson 
have impact on the $B$-decay observables \cite{neubert} and provide an unpleasant noise for
the studies of possible new physics effects (see, e.g., 
\cite{ciuchini2022,ciuchini2020,ciuchini2021,diego2021,matias2022,gubernari2022,hurth2022,stangl2022}).  

A number of theoretical analyses of nonfactorizable (NF) charming loops in FCNC $B$-decays have 
been published: In \cite{voloshin}, an effective gluon-photon local operator describing the charm-quark loop was  
calculated as an expansion in inverse charm-quark mass $m_c$ and applied to inclusive $B\to X_s\gamma$ decays
(see also \cite{ligeti,buchalla}). In \cite{khod1997}, NF corrections in $B\to K^*\gamma$ using local
operator product expansion (OPE) have been studied. NF corrections induced by the {\it local} photon-gluon operator 
were calculated in \cite{zwicky1,zwicky2} in terms of the light-cone (LC) 3-particle antiquark-quark-gluon
Bethe-Salpeter amplitude (3BS) of $K^*$-meson \cite{braun,ball1,ball2} with two field operators having equal coordinates,  
$\langle 0| \bar s(0)G_{\mu\nu}(0) u(x)|K^*(p)\rangle$, $x^2=0$. However, local OPE for the charm-quark loop in FCNC $B$ decays
leads to a power series in $\Lambda_{\rm QCD} m_b/m_c^2$; numerically this parameter is close to one.
To sum up $O(\Lambda_{\rm QCD} m_b/m_c^2)^n$ corrections, Ref.~\cite{hidr} obtained a {\it nonlocal} photon-gluon
operator describing the charm-quark loop and evaluated its effect making use of 3BS of the $B$-meson
in a {\it collinear} LC configuration $\langle 0| \bar s(x)G_{\mu\nu}(ux) b(0)|B(p)\rangle$, $x^2=0$ \cite{japan,Braun2017}. 
This approximation was later used for the analysis of other FCNC $B$-decays \cite{gubernari2020}.

The collinear LC configuration was known to provide the dominant 3BS contribution to meson tree-level 
form factors \cite{braun1994,offen2007}, in particular, to form factors of semileptonic (SL) $B$-decay
induced by $b\to u$ weak charged current (CC). It was tempting to use the collinear 3BS
for the description of FCNC $B$-decays. However, the 3BS contribution to the CC $B$-decay and to the FCNC
$B$-decay have a qualitative difference. To demonstrate this difference, let us consider the $B$-decay
in the $B$-meson rest frame.
In CC $B$-decays, the $b$-quark emits a fast light $u$-quark which is later hit by a soft gluon and thus
keeps moving in the {\it same} space direction. In the case of charming loops in FCNC $B$-decays,
a fast light $s$-quark and a pair of fast $c$-quarks emitted by the $b$-quark move in the {\it opposite} space
directions. In \cite{mk2018,m2019,m2022} it was demonstrated that the FCNC $B$-decay amplitude 
is dominated not by the collinear configuration of 3BS, but rather by a {\it double-collinear}
configuration $\langle 0| \bar q(x)G_{\mu\nu}(x') b(0)|B(p)\rangle$,
$x^2=0$, $x'^2=0$, but $(x-x')^2\ne 0$. The first application of a double-collinear
3BS to FCNC $B$-decays was presented in \cite{wang2022}. 

The noncollinear configuration of the $B$-meson 3BS leads to the appearance of new
Lorentz structures of the type $(x_\mu p_\nu-x_\mu p_\nu)/(xp)$ and $(x'_\mu p_\nu-x'_\mu p_\nu)/(x'p)$.
These Lorentz structures contain kinematical singularities at $1/(xp)$ and $1/(x'p)$ which should not
be singularities of the physical 3BS. This requirement and the continuity of the 3BS at the point
$x^2=0$, $x'^2=0$, $xp=0$, and $x'p=0$ yield constraints on the corresponding distribution amplitudes (DAs).
Moreover, these constraints turn out to be valid also for those DAs which appear in the collinear 3BS.
To avoid complications related to the spinorial structure of the 3BS in QCD, Sec.~\ref{Sect:2}
demonstrates the derivation of these constraints in the case of field theory with scalar ``quarks''.
A generalization to QCD is straightforward. 

In Sec.~\ref{Sect:3} we analyze Feynman diagrams of the type corresponding to charming loops in FCNC $B$-decays
(i.e., those diagrams in which the heavy field hits the middle point of the line along which light degrees of
freedom propagate; hereafter referred to as FCNC-type diagrams) and show that
to the leading order in the heavy-quark (HQ) expansion, the amplitude is given by the convolution of the hard kernel
composed of propagators of the light degrees of freedom and the $B$-meson 3BS in the double-collinear
configuration. We also derive the exact expression for the $B$-decay amplitudes of FCNC-type using
the $\alpha$-representation and show that the exact amplitudes and the double-collinear approximation
for these amplitudes coincide in the heavy-quark limit. 

%\newpage
\section{\label{Sect:2}Properties of the 3BS wave function of the $B$-meson}
Leaving aside technical details related to the spins of the constituent fields and of the bound state,
and considering heavy $B$-meson as bound state of scalar fields, one of which, $\phi_b(x)$, is heavy, 
the 3BS of the $B$-meson may be parametrized as follows (\cite{m2022} and refs therein): 
\begin{eqnarray}
  \langle 0|\phi(x)\phi_b(0)\phi'(x')|B(p)\rangle&=&
  \int D(\omega,\omega')
  e^{-ipx\omega -ipx'\omega'}\Psi(\omega,\omega',x^2,x'^2,(x-x')^2),
\end{eqnarray}
where 
\begin{eqnarray}
D(\omega,\omega')\equiv d\omega\, d\omega'\,\theta(\omega)\theta(\omega')\theta(1-\omega-\omega').
\end{eqnarray}
For $\Psi(\omega,\omega',x^2,x'^2,(x-x')^2)$ one can write down Taylor expansion in its variables $x^2$, $x'^2$, $(x-x')^2$:
\begin{eqnarray}
  \Psi(\omega,\omega',x^2,x'^2,(x-x')^2)&=&\Psi_0(\omega,\omega')+x^2 \Psi_{12}(\omega,\omega')
  +x'^2\Psi_{23}(\omega,\omega')+(x-x')^2\Psi_{13}(\omega,\omega')+\dots, 
\end{eqnarray}
where $\dots$ stand for higher powers of $x^2,x'^2,(x-x')^2$.
The distribution amplitudes (DAs) $\Psi_i$ have support in the region $0<\omega$, $0<\omega'$, $\omega+\omega'<1$,
and peak at small values $(\omega,\omega')=O(\Lambda_{\rm QCD}/M_B)$,
reflecting the fact that the heavy $b$-quark carries almost the full momentum of the heavy $B$-meson,
whereas the light degrees of freedom carry its small fraction $O(\Lambda_{\rm QCD}/M_B)$.
(In practical model calculations, the DAs may be nonzero in a narrower region, e.g.
$0<\omega$, $0<\omega$, $\omega+\omega'<2\omega_0<1$.)

Interesting constraints on the DAs emerge in those cases when the 3BS has a nontrivial Lorentz structure.
We shall consider a more complicated 3BS, when one of the light scalar fields is replaced by
the gauge field, $\phi(x)\to G_{\mu\nu}(x)$:\footnote{In QCD, the analogous 3BS is
$\langle 0|\bar q(x')G_{\mu\nu}(x)b(0)|B(p)\rangle$. This amplitude is not gauge-invarinat as it contains
field operators at different locations. Gauge-invariant 3BS is constructed by introducing the Wilson lines as follows
$\langle 0|\bar q(x')U(x',x)G_{\mu\nu}(x)U(x,0))b(0)|B(p)\rangle$ with
$U(x,y)=P \exp(i\int_y^{x} A_\mu(z)dz_\mu)$.
}
\begin{eqnarray}
\langle 0|G_{\mu\nu}(x)\phi_b(0)\phi'(x')|B(p)\rangle. 
\end{eqnarray}
Our discussion will be technically rather simple but will allow a direct generalization to QCD,
in which case more Lorentz structures emerge. 

%%%%%%%%%%%%%%%%%%%%%%%%%%%%%%%%%%%%%%%%%
\subsection{Collinear 3BS of $B$-meson}
In the literature, much attention has been given to the collinear 3BS, where the coordinates
of the light degrees of freedom are proportional to each other \cite{offen2007}:
\begin{eqnarray}
\label{1}
\langle 0|G_{\mu\nu}(u x)\phi_b(0)\phi'(x)|B(p))\rangle=
\int D(\omega,\omega')\,
e^{-i(u \omega+\omega') p x} \left(\frac{x_\mu p_\nu-x_\nu p_\mu}{xp}\right)\left[X_A(\omega,\omega')+O(x^2)\right], \; 0<u<1.
\end{eqnarray}
For $x^2=0$ and $xp\to 0$, all components of the 4-vector $x_\mu$ vanish in the rest frame of the $B$-meson
($p_+=p_-$, $p_\perp=0$):
\begin{eqnarray}
 && xp=x_+p_-+x_-p_+-x_\perp p_\perp=0\Longrightarrow x_+=-x_-;\nonumber
  \\
 && x^2=2x_+x_--x_\perp^2=-2x_+^2-x_\perp^2=0\Longrightarrow x_+=-x_-=0, x_\perp=0.
  \nonumber
\end{eqnarray}
So, if (\ref{1}) is defined only for $x^2=0$, 
one can introduce the new variable $\tau$, $x_\mu=\tau n_\mu$, $n^2=0$ ($n_\mu$ lies, e.g., along the $(+)$ direction
of the light cone). Then the Lorentz structure $(x_\mu p_\nu-x_\nu p_\mu)/xp$ has no singularity at $\tau\to 0$ and
therefore the amplitude (\ref{1}) is finite at $xp\to 0$ \cite{Braun2017}.

However, if one considers the collinear 3BS (\ref{1}) also at $x^2\ne 0$, then the singularity at $xp\to 0$
in the Lorentz structure emerges. The 3BS (\ref{1}) as the function of two arguments $x^2$ and $xp$ should be continuous
and finite at $x^2=0$ and $xp=0$, so one has to require that the kinematical singularity of the Lorentz structure should
not be the singularity of the 3BS (\ref{1}).
Expanding the exponential in (\ref{1}) and requiring the absence of a singularity at
$xp\to 0$ in the r.h.s. of Eq.~(\ref{1}) leads to a well-known constraint [see Eq.~(5.5) of \cite{Braun2017}]: 
\begin{eqnarray}
\label{c1}
\int D(\omega,\omega') X_A(\omega,\omega')=0.
\end{eqnarray}
%Model DAs considered in the literature for the corresponding Lorentz structures \cite{Braun2017}
%indeed satisfy this relation.
%Eq.~(\ref{c1}) does not know anything about the value of $x^2$, so it seems natural to require that (\ref{c1})
%is fulfilled also on the light cone $x^2=0$.
%%%%%%%%%%%%%%%%%%%%%%%%%%%%%%%%%%%%%%%%%%%%%%%%%%%%%

\subsection{Generalization to a noncollinear kinematics}
For a proper description of the nonfactorizable charm in FCNC $B$-decay, 3BS of the $B$-meson in a
noncollinear configuration
is necessary (see \cite{mk2018,m2019,m2022,wang2022} and the demonstration in the next Section).
We therefore consider the generalization of the
amplitude (\ref{1}) and allow the gluon
field $G_{\mu\nu}(x)$ and the light-quark field $\phi'(x')$ to have arbitrary different non-collinear coordinates. 
Then a more general decomposition emerges:\footnote{We do not display here
further structures like $(x_\mu x'_\nu-x'_\mu x_\nu)/((xp)\,(x'p))$. The way one should properly treat the
corresponding DAs will become obvious.}
\begin{eqnarray}
\label{1nc}
\langle 0|G_{\mu\nu}(x)\phi_b(0)\phi'(x')|B(p)\rangle&=&
\int D(\omega,\omega')\, 
e^{-i\omega x p-i\omega' x' p}
\nonumber\\
&&\times
\bigg\{
\left(\frac{x_{\mu} p_\nu-x_{\nu} p_\mu}{x\, p}\right)X^{(1)}_A(\omega,\omega')+
\left(\frac{x'_{\mu} p_\nu-x'_{\nu} p_\mu}{x'\,p}\right)X^{(2)}_A(\omega,\omega')+\dots \bigg \}.
\end{eqnarray}
The amplitude (\ref{1nc}) contains factors $1/x p$ and $1/x'p$ in the Lorentz structures. 
If (\ref{1nc}) is defined for $x^2=0$ and $x'^2=0$ only, one can introduce
two new variables $\tau $ and $\tau'$ such that $x_\mu = \tau n_\mu$, $x'_\mu = \tau' n'_\mu$, $n^2=0$
and $n'^2=0$, but $n'n\ne 0$
(e.g. $n_\mu$ lies along the $(+)$-direction of the light cone, and $n'_\mu$ lies along the $(-)$-direction, see also
\cite{wang2022}). Then the Lorentz structure $(x_\mu p_\nu-x_\nu p_\mu)/xp$ has no singularity at $\tau\to 0$,
the Lorentz structure $(x'_\mu p_\nu-x'_\nu p_\mu)/x'p$ has no singularity at $\tau'\to 0$, 
and therefore the amplitude (\ref{1nc}) has no singularity at $xp\to 0$ and $x'p\to 0$.

However, the noncollinear 3BS (\ref{1nc}) should be a
regular continuous function of its arguments at the point $x^2=0$, $x'^2=0$, $xp=0$, and $x'p=0$
independently of the way one approaches this point. If one first takes the limit $xp\to 0$ and $x'p\to 0$ keeping
$x^2\ne 0 $ and $x'^2\ne 0$, then the singularities at $xp\to 0$ and $x'p\to 0$ in the Lorentz structures emerge. 
To compensate these singularities, the DAs should satisfy the following constraints [obtained by
expanding the exponential in (\ref{1nc})]
\begin{eqnarray}
\label{c1nc1}
\int\limits_0^{1-\omega'} d\omega X^{(1)}_A(\omega,\omega')=0 \quad \forall \,\omega'\quad {\rm and}\quad
\int\limits_0^{1-\omega} d\omega' X^{(2)}_A(\omega,\omega')=0 \quad \forall \,\omega.
\end{eqnarray}
For $X_A$ parametrizing the 3BS in the collinear limit, $x=u x'$, one finds the relation
\begin{eqnarray}
\label{XAsum}
X_A(\omega,\omega')=X^{(1)}_A(\omega,\omega')+X^{(2)}_A(\omega,\omega')
\end{eqnarray}
Obviously, the condition (\ref{c1}) follows from (\ref{c1nc1}). However, Eq.~(\ref{c1nc1}) 
shows that each of the parts of $X_A(\omega,\omega')$ should satisfy more restrictive constraints. 
%
%In the next Section we shall calculate the $B$-decay amplitude of FCNC type in terms of $X_A$ making use of the 
%$\alpha$-integration. It will turn out that unless the conditions (\ref{c1nc1}) and (\ref{c1nc2}) are satisfied,
%the $B$-decay amplitude has an unphysical imaginary part. 

For further use we introduce the primitives: 
\begin{eqnarray}
  \label{2nc1}
&\overline{X}^{(1)}_A(\omega,\omega')=\int\limits_0^{\omega} d\underline{\omega} {X}^{(1)}_A(\underline{\omega},\omega'),\qquad
& {X}^{(1)}_A(\omega,\omega')=\partial_{\omega}\overline{X}^{(1)}_A(\omega,\omega')\\
  \label{2nc2}
&\overline{X}^{(2)}_A(\omega,\omega')=\int\limits_0^{\omega'} d\underline{\omega}' {X}^{(2)}_A(\omega,\underline{\omega}'),\qquad
& {X}^{(2)}_A(\omega,\omega')=\partial_{\omega'}\overline{X}^{(2)}_A(\omega,\omega').  
\end{eqnarray}
By virtue of (\ref{c1nc1}), we obtain the following contraints on the primitives
\begin{eqnarray}
\label{c2nc1}
&&\overline{X}^{(1)}_A(\omega=0,\omega')=\overline{X}^{(1)}_A(\omega=1-\omega',\omega')=0 \quad \forall \omega',\\
\label{c2nc2}
&&\overline{X}^{(2)}_A(\omega,\omega'=0)=\overline{X}^{(2)}_A(\omega,\omega'=1-\omega)=0 \quad\, \forall \omega. 
\end{eqnarray}
These relations mean that {\it the primitives should vanish on the boundary of the DA's-support region.}

Performing parts integration in $\omega$ or $\omega'$, and taking into account that the surface terms vanish due to 
(\ref{c2nc1}) and (\ref{c2nc2}), we may rewrite (\ref{1nc}) in the form that does
not contain factors $1/xp$ and $1/x'p$:\footnote{Noteworthy, the functions of the type $X_A$ extensively used in the
literature \cite{Braun2017,hidr,gubernari2020} do not satisfy (\ref{c1nc1}).
In this case, the surface terms do not vanish and the 3BS given by Eqs.~(\ref{1nc}) and (\ref{1ncm}) are not
equivalent to each other.}  
\begin{eqnarray}
\label{1ncm}
\langle 0|G_{\mu\nu}(x) \phi_b(0)\phi'(x')|B(p)\rangle&=& i
\int D(\omega,\omega')\,e^{-i\omega x p-i\omega' x' p} 
\nonumber \\ 
&&\times
\bigg\{
\left(x_{\mu} p_\nu-x_{\nu} p_\mu\right)\overline{X}^{(1)}_A(\omega,\omega')
+\left(x'_{\mu} p_\nu-x'_{\nu} p_\mu\right)\overline{X}^{(2)}_A(\omega,\omega')
+\dots\bigg \}. 
\end{eqnarray}
%Noteworthy, the functions of the type $X_A$ extensively used in the
%literature \cite{Braun2017,hidr,gubernari2020} do not satisfy (\ref{c1nc1}) and (\ref{c1nc2}).
%In this case, the 3BS given by Eqs.~(\ref{1nc}) and (\ref{1ncm}) are not equivalent to each other. 

Let us summarize the material of this Section: We required that the noncollinear 3BS is a continuous function of its
arguments at $x^2=0$, $x'^2=0$, $xp=0$ and $x'p=0$ and has a finite value at this point. Then, the kinematical singularities
in the Lorentz structures at $xp\to 0$ and $x'p\to 0$ should be compensated by certain properties of the DAs: namely,
for the DA of the $X_A$-type, its primitives should vanish on the boundary of $X_A$-support region. 
This property also guarantees that two forms of the 3BS, Eqs.~(\ref{1nc}) and (\ref{1ncm}), 
are equivalent and may be safely applied to the calculation of the amplitudes of $B$-meson decays.
%Moreover, we show in the next Section that the FCNC-type $B$-decay amplitudes calculated via such DAs do not
%have unphysical imaginary parts. 
%%%%%%%%%%%%%%%%%%%%%%%%%%%%%%%%%%%%%%%%%%%%%%%%%%%%%%%%%%%%%
%\newpage
\section{\label{Sect:3}
$B$-decay in the kinematics of charming loops in FCNC decays}
In this Section we focus on 3BS contributions involving charming loops in FCNC
$B$-decays and show that to the leading order in the HQ expansion the amplitude
is given by the convolution of the hard kernel and the $B$-meson 3BS in the
noncollinear kinematical configuration. 

\subsection{The dominant contribution to the FCNC $B$-decay amplitude involving charming loops}
As noticed in \cite{m2022} (see Appendix A for details) the kinematics of FCNC $B$-decay amplitudes
involving charming loops is equivalent to the 3BS correction to the $B$-decay form factor with the
difference that the heavy field, $\phi_b(0)$, is in the middle of the quark line. 
%%%%%%%%%%%%%%%%%%%%%%%%%%%%%%%%%%%
\begin{figure}[b!]
  \begin{center}
  \includegraphics[height=3.1cm]{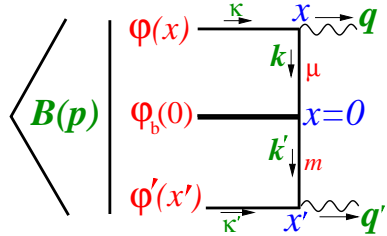}
  \caption{\label{Fig:1}
    The generic 3BS amplitude of the form factor topology. 
    If the field $\phi_b$ is heavy, while other fields $\phi$ and $\phi'$ are light,
    the diagram corresponds to the topology of charming loops in FCNC $B$-decay amplitude.}
\end{center}
\end{figure}
%%%%%%%%%%%%%%%%%%%%%%%%%%%%%%%%%%%
Placing the heavy-quark field at zero, $\phi_b(0)$, we have [Fig.~\ref{Fig:1}]:
\begin{eqnarray}
  \label{A0}
  A(p|q,q')=\int \frac{dx dx' dk dk'}{(\mu^2-k^2)(m^2-k'^2)}e^{i q x+i k x +i q' x'-i k' x'}
    \langle 0|\phi(x)\phi_b(0)\phi'(x')|B(p)\rangle, 
\end{eqnarray}
where $\phi$ and $\phi'$ are light degrees of freedom (in practical calculations, the gluon and the light quark). 
We are going to derive the leading-order behaviour of this amplitude. 

\subsubsection{The $x$-vertex}
Let us discuss the $x$-vertex and introduce $\kappa=k+q$, the momentum carried by the constituent field $\phi(x)$: 
\begin{eqnarray}
 \int \frac{dx d\kappa}{\mu^2-(\kappa-q)^2}e^{i \kappa x}
    \langle 0|\phi(x)...|B(p)\rangle.
\end{eqnarray}
We consider the case $q^2=q'^2=0$ and work in the rest frame of the $B$-meson and take the axes such that
momentum $q$ is along $(+)$-axis, whereas momentum $q'$ is along the $(-)$-axis.
[One can formulate this in a covariant form by introducing vectors $n_\mu$ and $\bar n_\mu$ \cite{Braun2017,wang2022}.] 

Due to the properties of the $B$-meson 3BS, the vector $\kappa$ is soft, i.e. all its components
are $\kappa_\mu\sim O(\Lambda_{\rm QCD})$.   
The component $q_+$ is large, $q_+\sim M_B$, and the propagator is highly virtual,
$\mu^2-2\kappa_-(\kappa_+-q_+)+\kappa_\perp^2\sim \Lambda_{\rm QCD}M_B$.

Let us expand the field operator $\phi(x)$ near $x=0$ (this would correspond to considering a tower of local operators of the
increasing dimension). The expansion in powers of $x_-$ and $x_\perp$ leads to a well behaved Taylor series as 
\begin{eqnarray}
  x_{-}e^{i\kappa_{+}x_{-}}
  \frac{1}{\mu^2-2\kappa_-(\kappa_+-q_+)+\kappa_\perp^2}
\to \frac{1}{\mu^2-2\kappa_-(\kappa_+-q_+)+\kappa_\perp^2}\partial_{\kappa_+}e^{i \kappa_+ x_-}\nonumber\\
\to e^{i \kappa_+ x_-}\partial_{\kappa_+}\left(\frac{1}{\mu^2-2\kappa_-(\kappa_+-q_+)+\kappa_\perp^2}\right)
\to e^{i \kappa_+ x_-}\frac{\kappa_-}{(\mu^2-2\kappa_-(\kappa_+-q_+)+\kappa_\perp^2)^2}. 
\end{eqnarray}
Since $\kappa_-=O(\Lambda_{\rm QCD})$ and the virtuality of the propagator is $O(\Lambda_{\rm QCD}M_B)$, 
any term $(x_-)^n$ is suppressed by a factor $(1/M_B)^n$ compared to the term $(x_-)^0$.
The same property holds for $(x_\perp)^n$.

However, for powers of the variable $x_+$ the situation is different: 
\begin{eqnarray}
  x_+ e^{i \kappa_- x_+}\frac{1}{\mu^2-2\kappa_-(\kappa_+-q_+)+\kappa_\perp^2}
  \to \frac{1}{\mu^2-2\kappa_-(\kappa_+-q_+)+\kappa_\perp^2}\partial_{\kappa_-}e^{i \kappa_- x_+}\nonumber\\
 \to \partial_{\kappa_-}\left(\frac{1}{\mu^2-2\kappa_-(\kappa_+-q_+)+\kappa_\perp^2}\right) e^{i \kappa_- x_+}
 \to e^{i \kappa_- x_+}\frac{q_+}{(\mu^2-2\kappa_-(\kappa_+-q_+)+\kappa_\perp^2)^2}.
\end{eqnarray}
Since $q_+\sim M_B$, all powers of $x_+^n$ in the expansion of $\phi(x_+)$ near $x_+=0$ have the same order of magnitude.
The Taylor expansion of $\phi(x_+)$ near $x_+=0$ leads to no hierarchy in the corresponding expansion of the $B$-decay amplitude,
and we need to keep the full $x_+$ dependence of the operator $\phi(x_+)$ on the light cone ($x^2=0$).
This result is valid for two phenomenologically interesting cases: (i) the charm-quark loop contribution
to the FCNC amplitude, 
$\mu^2\sim m_c^2=O(\Lambda_{\rm QCD}m_b)$, and (ii) the light-quark loop contribution to the FCNC amplitude, 
$\mu$ being the light-quark mass. 

So, the leading term of the expansion of the $B$-decay amplitude related to the $x$-vertex
corresponds to the expansion near $x_-=0,x_\perp=0$ and has the form 
\begin{eqnarray}
  \int dx_{+} dx_{-}dx_{\perp}  d\kappa_+ d\kappa_- d\kappa_\perp
  \frac{1}{\mu^2-2(\kappa_+-q_+)k_-+k_\perp^2}e^{i \kappa_+x_-+i \kappa_{-}x_+-ik_\perp x_\perp}
    \langle 0|...\phi(x_+)...|B(p)\rangle,
\end{eqnarray}
The $x_-$ and $x_\perp$ integrals here may be taken and lead to $\delta(\kappa_\perp)\delta(\kappa_+)$.
Integrating these $\delta$-functions, we obtain for the part of the amplitude related to the $x$-vertex 
(we denote $\tau=x_+$, and recall that $q$ has only the $(+)$-component): 
\begin{eqnarray}
  \int d\tau d\kappa_-\frac{1}{\mu^2 + 2 q_+ \kappa_-}e^{i \kappa_-\tau}
  \langle 0|\phi(\tau)...|B(p)\rangle.
\end{eqnarray}
\subsubsection{The $x'$-vertex \label{doublecollinear}}
We can perform a similar analysis of the $x'$ vertex. The crucial difference is that
now $q_-\sim M_B$ is the only nonzero component of the vector $q'$. The propagator has the form
$m^2-2\kappa'_-(\kappa'_--q_-)+{\kappa'_\perp}^2\sim \Lambda_{\rm QCD}M_B$. Obviously, we can perform Taylor expansion
of $\phi'(x')$ near $x'_+=0$ and $x'_\perp=0$ but have to keep its full dependence on the variable $x_-$.
Taking into account this property and denoting $\tau'=x'_-$, the dominant contribution of the $x'$-vertex reads
($q_-\sim M_B$ is the only nonzero component of the vector $q'$)
\begin{eqnarray}
  \int d\tau' d\kappa'_+\frac{1}{m^2 + 2 q'_- \kappa'_+}e^{i \kappa'_+\tau'}
  \langle 0|...\phi'(\tau')...|B(p)\rangle.
\end{eqnarray}
\subsubsection{The amplitude of the FCNC $B$-decay}
Making use of the leading contributions of the $x$- and $x'$-vertices 
we obtain the leading contribution to the $B$-decay amplitude in the form, see also \cite{wang2022}: 
\begin{eqnarray}
  \label{Apqfinal}
  A(p|q,q')=
  \int d\tau d\kappa_-\frac{1}{\mu^2 + 2 q_+ \kappa_-}e^{i \kappa_-\tau}
  \int d\tau' d\kappa'_+\frac{1}{m^2 + 2 q'_- \kappa'_+}e^{i \kappa'_+\tau'}
  \langle 0|\phi(\tau)\phi_b(0)\phi'(\tau')|B(p)\rangle.
\end{eqnarray}
This relation is the factorization theorem that represents the dominant contribution to the FCNC amplitude
as the convolution of the hard kernel composed from the propagators of the light degrees of freedom
and the 3BS in the kinematical configuration which may be called ``double collinear'':
the upper and the lower parts of the diagram are aligned along the different light-cone directions.

%%%%%%%%%%%%%%%%%%%%%%%%%%%%%%%%%%%%
\subsubsection{On the collinear light-cone 3DA in FCNC $B$-decay}
We would like to emphasize that the 3BS in a collinear light-cone configuration
\begin{eqnarray}
\label{coll3DA}
\langle 0|\phi(x)\phi_b(0)\phi'(\lambda x)|B(p)\rangle,\quad x^2=0, \quad 0<\lambda<1, 
\end{eqnarray}
is {\it not related} to the dominant contribution to the amplitude of FCNC $B$-decay.
So, we do not find justification of the statement of \cite{hidr} that the dominant contribution
to the FCNC amplitude may be calculated via the collinear light-cone 3DA of the $B$-meson (\ref{coll3DA}).

%%%%%%%%%%%%%%%%%%%%%%%%%%%%%%%%%%%  MULTI-BS %%%%%%%%%%%%%%%
\subsubsection{Multi-particle BS contributions to amplitudes of $B$-decays}
\begin{figure}[h!]
    \begin{center}
\includegraphics[height=4.6cm]{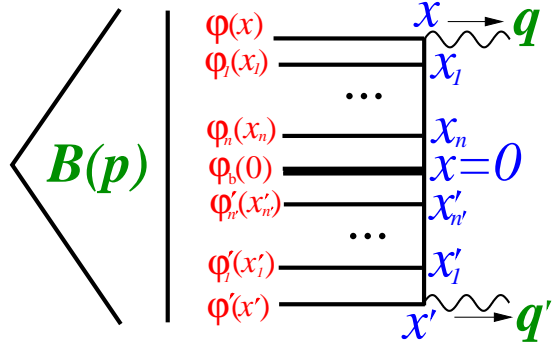}
\caption{\label{Fig:2}
An example of a multi-particle BS contribution to the $B$-decay amplitude.
The field $\phi_b$ is heavy, while all other fields, $\phi$, $\phi'$, \ldots, are light.
The dominant contribution comes from the double collinear LC configuration:
$x_1=u_1 x$, \ldots, $x_n=u_n x$, $0<u_n<\ldots <u_1<1$, $x^2=0$; 
$x'_1=u'_1 x'$, \ldots, $x'_{n'}=u_{n'} x'$, $0<u'_{n'}<\ldots <u'_1<1$, $x'^2=0$,
whereas $xx'\ne 0$,
i.e., the set of collinear field coordinates in the upper part of the diagram is not aligned with 
the set of collinear field coordinates in the lower part of the diagram.}
  \end{center}
\end{figure}

Our analysis may be generalized to other contributions to $B$-decay amplitudes
of the type shown in Fig.~\ref{Fig:2}.
The corresponding diagram involves a multi-particle BS of the $B$-meson: 
\begin{eqnarray}
  \langle 0|\phi(x)\phi_1(x_1)\ldots \phi_n(x_n) \phi_b(0)
  \phi'_{n'}(x'_{n'})\ldots\phi_1'(x_1')\phi'(\tau')|B(p)\rangle,   
\end{eqnarray}
where all fields except for $\phi_b$ are light. 
Combining the analysis presented above and the discussion of Sect.~3 of \cite{m2022},
one can easily prove by induction that the dominant contribution to the $B$-decay amplitude
comes from the double-collinear light-cone configuration 
[$a_\mu=(a_+,a_-,a_\perp)$]:
\begin{eqnarray}
%\begin{multiline}
& x=(\tau,0,0),    \quad & x_1=(\tau u_1,0,0),\quad \,\,\ldots,\quad     x_n=(\tau u_n,0,0),   \qquad 0<u_n<\ldots<u_1<1, \nonumber\\
&  x'=(0,\tau',0), \quad & x'_1=(0,\tau' u'_1,0),\quad \ldots,\quad x'_{n'}=(0,\tau' u'_{n'},0), \,\quad 0<u'_{n'}<\ldots<u'_1<1.
%\end{multiline}
\end{eqnarray}
The coordinates $x,x_1,\ldots,x_n$ are ordered and lie on the $(+)$-axis of the LC, 
and the coordinates $x',x'_1,\ldots,x'_{n'}$ are ordered and lie on the $(-)$-axis of the LC.
That is why we refer to this configuration as to the double collinear light-cone configuration. 

\subsubsection{Collinear vs Double-collinear 3DAs}
Here we would like to emphasize the differences between the collinear and the double-collinear 3DAs. 
Let us get back to Eq.~(\ref{1nc}). 
%and consider the following contribution to the $B$-meson 3BS:
%\begin{eqnarray}
%\label{AppBXA}
%\langle 0|G_{\mu\nu}(x) \phi_b(0)\phi(x')|B(p)\rangle&=&\int D(\omega,\omega')
%e^{-i\omega x p-i\omega' x' p}\nonumber\\
%&&\times\left\{
%\left(\frac{x_{\mu} p_\nu-x_{\nu} p_\mu}{x\,p}\right)X^{(1)}_A(\omega,\omega')+
%\left(\frac{x'_{\mu} p_\nu-x'_{\nu} p_\mu}{x'\,p}\right) X^{(2)}_A(\omega,\omega')\right\}.
%\end{eqnarray}
Again introduce LC variables %[$a_\mu=(a_+,a_-,a_\perp)$, $a^2=2a_+a_--a_\perp^2$]
in the $B$-meson rest frame and consider two vectors $n_\mu=(\sqrt2,0,0)$ and $n'_\mu=(0,\sqrt2,0)$ such that 
$n^2=0$, $n'^2=0$, $nn'=2$, $v_\mu=\frac12(n_\mu+n'_\mu)$, $nv=n'v=1$.
  
\noindent 
$\bullet$ In a collinear LC configuration, $x^2=0$, $x'^2=0$, $x'_\mu = u x_\mu$, 
we can choose $x_\mu$ and $x'_\mu$ along $n_\mu$: $x_\mu=\tau n_\mu$,  $x'_\mu=\tau' n'_\mu$.
Then two Lorentz structures in (\ref{1nc}) reduce to one Lorentz structure (we use relation $v=(n+n')/2$):
\begin{eqnarray}
\label{AppBXA1}
\langle 0|G_{\mu\nu}(n \tau) \phi_b(0)\phi(n\tau')|B(p)\rangle=
\int D(\omega,\omega')
e^{-i M_B(\omega\tau+\omega'\tau')} 
\frac{1}{2}\left(n_{\mu} n'_\nu-n_{\nu} n'_\mu \right)\left(X^{(1)}_A+X^{(2)}_A\right), 
\end{eqnarray}
where $X_A\equiv X^{(1)}_A+X^{(2)}_A$ is the collinear LC 3DA.
%Assuming that the 3DAs are proportional to each other, we obtain 
%\begin{eqnarray}
%\label{AppBXA1a}
%X^{(x)}_A=g_x X_A,\quad X^{(x')}_A=g_{x'} X_A,\quad g_x+g_{x'}=1. 
%\end{eqnarray}

\noindent $\bullet$
We now turn to the double-collinear LC configuration. 
Since the 3BS (\ref{1nc}) is antisymmetric in $\mu\leftrightarrow\nu$, 
both Lorentz structures in (\ref{1nc}) are again reduced to the same Lorentz structure as in (\ref{AppBXA1}) 
with however a different 3DA:
\begin{eqnarray}
\label{AppBXA2}
\langle 0|G_{\mu\nu}(n \tau) \phi_b(0)\phi(n'\tau')|B(p)\rangle=
\int D(\omega,\omega')
e^{-i M_B(\omega\tau+\omega'\tau'} 
\frac{1}{2}\left(n_{\mu} n'_\nu-n_{\nu} n'_\mu\right)\left(X^{(1)}_A-X^{(2)}_A\right).
\end{eqnarray}
So, we conclude: Although the 3BS of Eq.~(\ref{1nc}) both in the collinear and the double-collinear
configurations contain one and the same Lorentz structure, the corresponding 3DAs in the collinear and the 
double-collinear configurations are different and in general are independent of each other, see also
\cite{wang2022}. 

%%%%%%%%%%%%%%%%%%%%%%%%%%%%%%%%%%%%%%%%%%%%%%%%%%%%%%%%%%%%%%%%%%%%%%%%%%%%%%%
\subsection{Gauge field as one of the light fields: double-collinear approximation}
Let us now turn to the situation with more Lorentz structures.
We have to take into account that the kinematics is double-collinear
and the new Lorentz structures and new DAs emerge compared to collinear kinematics
considered in \cite{Braun2017}. 
Still, some features established for the collinear 3DA survive also in this more complicated case.
For instance, let us consider the Lorentz structures containing $xp$ and $x'p$ in the denominator 
\begin{eqnarray}
\frac{x_{\mu} p_\nu -x_{\nu} p_\mu}{x\,p}\quad {\rm and} \quad \frac{x'_{\mu} p_\nu -x'_{\nu} p_\mu}{x'p}.
\end{eqnarray}
If we keep the ``large'' components (i.e. $x_{\mu}$ along the $(+)$-axis and
$x'_{\mu}$ along the $(-)$-axis), then powers of $\tau$ and $\tau'$ in the numerator
and the denominator cancel. In the end, the $\tau$- and $\tau'$-integrations may be easily performed.

To be more specific, we consider the case when the light field $\phi(x)$ is replaced
by the vector gauge field $G_{\mu\nu}(x)$ and discuss one structure in the $B$-meson 3BS 
\begin{eqnarray}
\label{XA}
\langle 0|G_{\mu\nu}(x) \phi_b(0)\phi(x')|B(p)\rangle=
\int D(\omega,\omega')
e^{-i\omega x p-i\omega' x' p} 
\left(\frac{x_{\mu} p_\nu-x_{\nu} p_\mu}{x\,p}\right)X_A(\omega,\omega')
\end{eqnarray}
The $B$-decay amplitude induced by this 3BS is anti-symmetric in two indiices $\mu$ and $\nu$
and thus contains one form factor: 
\begin{eqnarray}
  \label{Apqfinal2}
A_{\mu\nu}(p|q,q')&=&(q_\mu q'_\nu-q_\nu q'_\mu)F(q^2,q'^2)\nonumber\\
                 &=&\int \frac{dx dx' d\kappa d\kappa'}{(\mu^2-(\kappa-q)^2)(m^2-(\kappa'-q')^2)}e^{i\kappa x +i \kappa'x'}
  \langle 0|G_{\mu\nu}(x) \phi_b(0)\phi'(x')|B(p)\rangle, 
\end{eqnarray}
Taking into account the derivation of the previous subsection, the leading contribution may be written as
\begin{eqnarray}
A_{\mu\nu}(p|q,q')&=&\int d\tau d\kappa_-\frac{1}{\mu^2 + 2 q_+ \kappa_-}e^{i \kappa_-\tau}
  \int d\tau' d\kappa'_+\frac{1}{m^2 + 2 q'_- \kappa'_+}e^{i \kappa'_+\tau'}
  \frac{x_{\mu} p_\nu -x_{\nu} p_\mu}{x\,p}\nonumber\\
 &&\times \int D(\omega,\omega')\, X_A(\omega,\omega')e^{-i \omega  p_-\tau-i\omega' p_+\tau'}.
\end{eqnarray}
Obviously, the main contribution to the amplitude comes from the following regions: 
\begin{itemize}
  \item
In the upper part of the diagram, $x_{\mu}$ has nonzero $(+)$-component, $q_\mu$ has nonzero $(+)$-component,
$\kappa_{\mu}$ has nonzero $(-)$-component, and $x p=\tau p_+$. 
\item
In the lower part of the diagram, the situation is opposite:
$x'_{\mu}$ has nonzero $(-)$-component, $q'_\mu$ has nonzero $(-)$-component,
$\kappa'_{\mu}$ has nonzero $(+)$-component, and $x'p=\tau' p_-$.
\end{itemize}
Essentially, we have two independent one-dimensional configurations. These configurations talk to each other via the
$\omega$ and $\omega'$ dependence of the 3DA $X_A(\omega,\omega')$. 

In the double-collinear configuration, the 4-vector $x$ has only one nonzero component, $x_+$, and the combination
$x_{\mu} p_\nu -x_{\nu} p_\mu$ is linear in $x_+$. Therefore the combination
\begin{eqnarray}
\frac{x_{\mu} p_\nu -x_{\nu} p_\mu}{x\,p}
\end{eqnarray}
is non-singular for $x_+\to 0$. [The same property holds e.g. for the structure ${(x_{\mu} \gamma_\nu -x_{\nu} \gamma_\mu)}/{x\,p}$
for spinor fields.]

Taking into account the nonzero components, for calculating the form factor $F(q^2,q'^2)$ we may set $\mu=(+)$ and $\nu=(-)$,
or $\mu=(-)$ and $\nu=(+)$. In the double-collinear configuration, we find 
\begin{eqnarray}
&&\frac{x_{\mu} p_\nu -x_{\nu} p_\mu}{xp}\to 1\qquad \mbox{ for } \mu=(+), \nu=(-), \mbox{ and}\nonumber\\
&&\frac{x_{\mu} p_\nu -x_{\nu} p_\mu}{xp}\to -1\qquad\mbox{ for } \mu=(-), \nu=(+).
\end{eqnarray}
In the end, we obtain the same result as for the scalar form factor of the previous Section: 
\begin{eqnarray}
  \label{Fdc}
  q'q\,F(q^2,q'^2) = \int d\tau d\kappa_-\frac{1}{\mu^2 + 2 q_+ \kappa_-}e^{i \kappa_-\tau}
  \int d\tau' d\kappa'_+\frac{1}{m^2 + 2 q'_- \kappa'_+}e^{i \kappa'_+\tau'}
\int D(\omega,\omega')\,e^{-i \omega p_-\tau -i\omega' p_+\tau'}\,X_A(\omega,\omega').\nonumber\\
\end{eqnarray}

%\newpage
%%%%%%%%%%%%%%%%%%%%%%%%%%%%%%%%%%%%%%%%%%%
\subsection{Gauge field as one of the light fields: the $\alpha$-representation}
Another possibility to handle the $1/xp$ factor without making approximations and without performing parts integration
is to make use of the $\alpha$ -representation
\begin{eqnarray}
\frac{1}{ixp}=\int_0^\infty e^{- i a px -\epsilon a} d a.
\end{eqnarray}
This representation leads to a shift of the expressions containing $\omega$ in the denominator of the quark propagator
and enables us to take explicitly the $\alpha$-integration in the amplitude (\ref{Apqfinal2}). So, we do not need
to make the parts integration in $\omega$. In this way we could avoid making approximations and considering the collinear kinematics. 

Taking into account that the quadratic form in the denominator is the quark propagator, we reconstruct the imaginary 
part as follows
\begin{eqnarray}
  \label{alpha}
  \int\limits_0^\infty\frac{d a}{(a-a_+)(a-a_-)- i 0}=\frac{1}{a_+-a-}\log\left({-\frac{a_-}{a_+}}\right)
  +i\pi\frac{1}{a_+-a-}, \quad  a_-<0, \quad a_+>0\quad  \mbox{for } 0<\omega <1. 
\end{eqnarray}
Let us discuss the contribution to the form factor from the ``upper part of the diagram''
including the propagator of the quark with mass $\mu$; the ``lower'' part of the diagram is treated precisely the same way.
Important is that we do not need to make any approximations and the calculation is fully covariant.
By virtue of the $\alpha$-representation, the contribution of the 3DA structure (\ref{XA})
to the form factor (\ref{Apqfinal}) takes the form (we explicitly write only the upper part of the diagram
containing the vertex $x$ that emits the momentum $q$, the quark propagator $1/(\mu^2-k^2-i 0)$ and
the relevant part of the $B$-meson 3BS): 
\begin{eqnarray}
  \label{A1}
  A_{\mu\nu}(p|q,q')=
  \int dx dk e^{iqx+ikx-i\omega px}\frac{x_\mu p_\nu-x_\nu p_\mu}{ixp}\frac{1}{\mu^2-k^2-i 0}d\omega X_A(\omega,\omega')\{\ldots\}.
\end{eqnarray}
We substitute $1/px$ as the $\alpha$-integral and representing the $x_\beta$ as the $q_\beta$-derivative of $\exp(iq x)$.
This $q_\beta$ derivative may be taken out of the integral bringing us to the following expression: 
\begin{eqnarray}
  \label{A2}
A_{\mu\nu}(p|q,q')=\left(p_\nu \frac{\partial}{\partial q_\mu}-p_\mu \frac{\partial}{\partial q_\nu}\right)
\int\limits_0^\infty da\, dx\, dk\,e^{iqx+ikx-i\omega px-ia px-\epsilon a}\frac{1}{\mu^2-k^2-i 0}d\omega X_A(\omega,\omega'\{\ldots\}
\end{eqnarray}
Taking $x$- and $k$-integrations yields
\begin{eqnarray}
  \label{A3}
A_{\mu\nu}(p|q,q')=\left(p_\nu \frac{\partial}{\partial q_\mu}-p_\mu \frac{\partial}{\partial q_\nu}\right)
\int\limits_0^\infty da e^{-\epsilon a}\frac{1}{\mu^2-(q-p(a+\omega))^2-i0}d\omega X_A(\omega,\omega')\{\ldots\}
\end{eqnarray}
The $q$-derivative may be written as
\begin{eqnarray}
  \frac{\partial}{\partial q_\beta}\left(\frac{1}{\mu^2-(q-p(a+\omega))^2-i0}\right)
  &=& -2\frac{\partial}{\partial \mu^2 }\left(\frac{1}{\mu^2-(q-p(a+\omega))^2-i0\cdots}\right)
  \left(q_\beta-p_\beta(a+\omega) \right)\nonumber\\
  &\to&
  -2q_\beta\frac{\partial}{\partial \mu^2 }\left(\frac{1}{\mu^2-(q-p(a+\omega))^2-i0\cdots}\right)
\end{eqnarray}
since the term proportional to $p_\beta$ does not contribute to the anti-symmetric amplitude $A_{\mu\nu}$.

\noindent
The next steps in our handling of Eq.~(\ref{A3}) are as follows:\\
The $a$-integral converges so we may take the limit $\epsilon\to 0$ and set $\exp(-\epsilon a)\to 1$.\\
Isolating the factor $M_B^2$ in the denominator, the $a$-integral in Eq.~(\ref{A3})
is equal to (\ref{alpha}) with 
\begin{eqnarray}
  \label{a+-}
a_+=\frac{-1+\sqrt{1+4\bar\mu^2}+2\omega}{2}>0, \quad 
a_-=\frac{-1-\sqrt{1+4\bar\mu^2}+2\omega}{2}<0, \quad {\rm for}\quad 0<\omega <1,  
\end{eqnarray}
where $\bar\mu^2\equiv \mu^2/M_B^2$ and $a_+-a_-=\sqrt{1+4\bar\mu^2}$. For further use we denote the
real part of (\ref{alpha}) as $K(\omega,\bar\mu^2)$: 
\begin{eqnarray}
  K(\omega,\bar\mu^2)=\frac{1}{a_+-a-}\log\left({-\frac{a_-}{a_+}}\right),
\end{eqnarray}
with $a_{\pm}$ given by (\ref{a+-}). Notice a useful relation
\begin{eqnarray}
\partial_\omega K(\omega,\bar\mu^2)=\frac{1}{\bar\mu^2+\omega(1-\omega)}
\end{eqnarray}
%\newpage
Adding the contribution of the ``lower'' part of the diagram yields
\begin{eqnarray}
  \label{Aqpfinal2}
  A_{\mu\nu}(p|q,q')\sim \left(q'_\nu q_\nu-q_\nu q'_\mu\right) \frac{\partial}{\partial \bar\mu^2}F_s(\bar\mu^2,\bar m^2),
\end{eqnarray}
with (recall that we consider $q^2=q'^2=0$) 
\begin{eqnarray}
  \label{Fs}
F_s(\bar \mu^2,\bar m^2)&=&\int D(\omega,\omega')\, X_A(\omega,\omega')K(\omega,\bar\mu^2)
\frac{1}{\bar m^2+\omega' (1+\omega')}\nonumber\\
&&+i\pi \frac{1}{\sqrt{1+4\bar\mu^2}}
\int D(\omega,\omega')\, X_A(\omega,\omega')\frac{1}{\bar m^2+\omega'(1+\omega')}. 
\end{eqnarray}
Here $\bar m^2\equiv m^2/M_B^2$.

To compare the amplitude in the double-collinear approximation, Eq.~(\ref{Fdc}),
and the exact amplitude, Eq.~(\ref{Aqpfinal2}),
in the HQ limit ($M_B\to\infty$, $m$ and $\mu$ fixed, $\omega\sim \Lambda_{\rm QCD}/M_B$),
we make use of the following expansions
\begin{eqnarray}
  \label{expansions1}
  -\partial_{\bar\mu^2}K(\omega,\bar\mu^2)&=& \frac{1}{\omega}-(1+2\log(\omega)-2 i\pi)-3\omega+O(\omega^2),\\
  \label{expansions2}
  \frac{1}{\bar\mu^2+\omega(1-\omega)}&=& \frac{1}{\omega} + 1+\omega+O(\omega^2).
\end{eqnarray}
Taking into account that any 3DA is peaked in region ($\omega,\omega') \sim \Lambda_{\rm QCD}/m_b$ , 
we observe the following important features of the amplitude in the HQ limit:
\begin{itemize}
\item[(a)]
The amplitude in the ``double-collinear'' approximation, Eqs.~(\ref{Fdc}) and (\ref{expansions2})
and the exact amplitude, Eqs.~(\ref{Aqpfinal2}) and (\ref{expansions1}),
have one and the same leading behaviour in the HQ limit.
\item[(b)]
  The imaginary part of the exact amplitude, see Eqs.~(\ref{Aqpfinal2}) and (\ref{Fs}),
  is parametrically suppressed compared to its real part. The amplitude in the ``double-collinear''
  approximation (\ref{Fdc}) does not have any imaginary part.
\item[(c)] 
  The ``strong'' imaginary part of the exact $B\to\gamma\gamma$ amplitude, Eq.~ (\ref{Fs}), 
  gained due to soft gluon interactions, is unphysical as there are no appropriate hadron
  intermediate states which might lead to the appearance
  of such imaginary part at $q^2=0$ and $q'^2=0$. Noteworthy, the imaginary part vanishes identically if $X_A$ satisfies
  the constraint (\ref{c1nc1}); the constraint on $X_A$ given by Eq.~(\ref{c1}) is not sufficient to guarantee
  the absence of the unphysical imaginary part. 
\end{itemize} 
%%%%%%%%%%%%%%%%%%%%%%%%%%%%%%%%%%%%%%%%%%%%%%%%%
%
% CONCLUSIONS
%
%%%%%%%%%%%%%%%%%%%%%%%%%%%%%%%%%%%%%%%%%%%%%%%%%
%\newpage
\section{Conclusions}
We studied a generic $B$-decay amplitude of the FCNC-type --- i.e., an amplitude given by diagrams in which the heavy
field hits the middle point of the line along which light degrees of freedom propagate --- and obtained the following results: 

\vspace{.5cm}
\noindent
(i) As already demonstrated in the literature \cite{mk2018,m2019,m2022}, the leading contribution to the amplitude of a
$B$-decay of FCNC-type, $B\to j(q)j'(q')$, is given by the convolution formula of the hard kernel composed of propagators
of the light degrees of freedom and the 3BS of the $B$-meson 
\begin{eqnarray}
\langle 0|\phi(x)\phi_b(0)\phi'(x')|B(p)\rangle 
\end{eqnarray}
in the following configuration:
\begin{eqnarray}
x^2=0; \qquad x'^2=0, \qquad xx'\ne 0. 
\end{eqnarray}
We have now formulated this result as a factorization theorem by a direct analysis of Feynman diagrams:
Considering the FCNC-type $B$-decay into two real photons ($q^2=q'^2=0$) in the rest frame of
the $B$-meson and choosing the momenta $q$ and $q'$ along the $(+)$ and $(-)$ axes of the light cone,
respectively, we have shown that the dominant contribution to the amplitude in the heavy-quark limit
comes from the ``double-collinear'' configuration 
\begin{eqnarray}
\label{doublecollinear3BS} 
\langle 0|\phi(x_+)\phi_b(0)\phi'(x'_-)|B(p)\rangle.  
\end{eqnarray}
Eq.~(\ref{Apqfinal}) represents the factorization formula for the amplitudes of FCNC-type; corrections to the contribution
given by Eq.~(\ref{Apqfinal}) are suppressed by powers of the heavy-quark mass. 

\vspace{.5cm}
\noindent
(ii) The $B$-meson 3BS in a collinear LC configuration
\begin{eqnarray}
  \label{collinear3BS}
\langle 0|\phi(x)\phi_b(0)\phi'(\lambda x)|B(p)\rangle, \qquad x^2=0, \qquad 0<\lambda <1,
\end{eqnarray}
{\it does not appear} in the convolution formula (\ref{Apqfinal}) for the leading
3BS contribution to the FCNC $B$-decay amplitude. This means that the calculation of the FCNC $B$-decay amplitude
as the convolution of the hard kernel and the collinear 3BS as is done e.g. in \cite{hidr} does not provide
a well-defined starting point: corrections to this approximation are not suppressed by any large parameter.

This makes an essential difference between the 3DA contributions to charming loops in FCNC $B$-decays
and to SL $B$-decays: 
The 3DA contribution to the SL form factor factorizes into the convolution of a hard kernel and the collinear LC
3BS \cite{offen2007}, corrections to this configuration being suppressed by $1/m_b$.

\vspace{.5cm}
\noindent
(iii) The generic 3BS involves more Lorentz structures and 3DAs than the collinear 3BS.
In particular, it involves structures of the type
%$D(\omega,\omega')=d\omega d\omega'\theta(\omega)\theta(\omega)\theta(1-\omega-\omega')$ 
\begin{eqnarray}
\label{3BSXA} 
\langle 0|G_{\mu\nu}(x)\phi_b(0)\phi'(x')|B(p)\rangle&=&
\int D(\omega,\omega')\;
e^{-ipx\omega-ipx'\omega'}\nonumber\\
&&\times\left\{
\left(\frac{p_\mu x_\nu-p_\nu x_\mu}{xp} \right) X^{(1)}_A(\omega, \omega')
+
\left(\frac{p_\mu x'_\nu-p_\nu x'_\mu}{x'p}\right) X^{(2)}_A(\omega, \omega')
\right\}
\end{eqnarray}
with $D(\omega,\omega')=d\omega d\omega'\theta(\omega)\theta(\omega)\theta(1-\omega-\omega')$.
These Lorentz structures contain kinematical singularities in $1/xp$ or $1/x'p$. Such singularities are
unphysical. Requiring that the 3DA is a continuous and finite function of its variables
at $x^2=0$, $x'^2=0$, $xp=0$ and $x'p=0$, we conclude that the kinematical singularities 
should vanish due to the properties of the 3DAs $X^{(1)}_A$ and $X^{(2)}_A$, leading to the following constraints:
\begin{eqnarray}
%  \label{cXa1}
  \int\limits_0^{1-\omega} d\omega X^{(1)}_A(\omega,\omega')=0\quad \forall \omega'; \qquad
%  \label{cXa2}
  \int\limits_0^{1-\omega'} d\omega' X^{(2)}_A(\omega,\omega')=0\quad \forall \omega.
\end{eqnarray}
In other words, the primitives
$\overline{X}^{(1)}_A(\omega,\omega')$ and $\overline{X}^{(2)}_A(\omega,\omega')$ [Eqs.~(\ref{2nc1}) and (\ref{2nc2})]
should vanish on the boundaies of the $X_A$-suppport area:
\begin{eqnarray}
&&  \overline{X}^{(1)}_A(\omega=0,\omega')=\overline{X}^{(1)}_A(1-\omega',\omega')=0 \quad \forall \omega',\nonumber\\
&&  \overline{X}^{(2)}_A(\omega,\omega'=0)=\overline{X}^{(2)}_A(\omega,1-\omega)=0 \quad \forall \omega.
\end{eqnarray}
Constraints of this type emerge for all 3DAs which parametrize the Lorents structures
$x_\mu/xp$, $x'_\mu/x'p$, $x_\mu x_\nu/(xp)^2$, $x_\mu x'_\nu/(xp)(x'p)$ etc in the 3BS of the $B$-meson. 
These constraints should be taken into account when building models for the 3DAs.  
%In general, they lead to a strong reduction of the contributions of such 3DAs to the amplitudes of FCNC $B$-decays.

\vspace{.5cm}
\noindent
(iv) The $B$-meson 3BS of the $X_A$-type of Eq.~(\ref{3BSXA}) in the collinear LC configuration $x_\mu=\tau n_\mu$,
$x'_\mu=\tau' n_\mu$, $n^2=0$, Eq.~(\ref{AppBXA1}),
and in the double-collinear LC configuration, $x_\mu=\tau n_\mu$,
$x'_\mu=\tau' n'_\mu$, $n^2=n'^2=0$, $n'n=2$, Eq.~(\ref{AppBXA2}), 
may be parametrized via one and the same Lorentz structure $n_\mu n'_\nu-n_\nu n'_\mu$. However, the corresponding
3DAs are different and in general independent of each other: 
the collinear LC 3DA is equal to $X^{(1)}_A+X^{(2)}_A$, whereas the double-collinear LC 3DA is equal to $X^{(1)}_A-X^{(2)}_A$.

\vspace{1cm}\noindent 
{\it Acknowledgments.}
I am grateful to I.~Belov, A.~Berezhnoy, M.~Ferre, E.~Kou, O.~Nachtmann, and H.~Sazdjian for 
discussions and to Q.~Qin, Y.-L.~Shen, C.~Wang and Y.-M.~Wang for comments on their paper \cite{wang2022}.
I am particularly indebted to Yu-Ming Wang for his illuminating remarks. 
%The research was carried out within the framework of the program ``Particle Physics and Cosmology'' 
%of the National Center for Physics and Mathematics.
\newpage

\appendix
\section{Nonfactorizable charming loop in FCNC $B$-decay amplitude}
\begin{figure}[ht!]
 \begin{center}   
 \includegraphics[height=4cm]{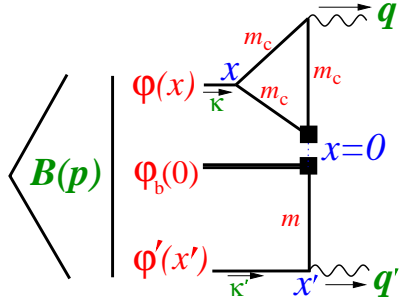}
 \caption{\label{Fig:3} Diagram describing $A_{\rm FCNC}$, a nonfactorizable charming-loop corrections
   to the amplitude of FCNC $B$-decay.}
 \end{center}
  \end{figure}

Nonfactorizable charm-loop contribution to the amplitude of FCNC decay is given by the diagram
of Fig.~\ref{Fig:3}. The corresponding analytic expression reads \cite{m2022}
\begin{eqnarray}
  \label{AFCNC}
A_{\rm FCNC}(p|q,q')=
\frac{G_F}{\sqrt2}\int d\kappa' dx' e^{i\kappa'x'}\frac{1}{m^2-(\kappa'-q')^2-i0}
d\kappa dx e^{i\kappa x} \Gamma_{cc}(\kappa,q)\langle 0 |\phi(x)\phi_b(0)\phi'(x')|B(p)\rangle,
\end{eqnarray}
where the expression for the charm-quark loop in the case of scalar ``quarks'' has the form
\begin{eqnarray}
  \label{charmloop}
\Gamma_{cc}(\kappa,q)=
\frac{1}{8\pi^2}\int\limits_0^1 du \int\limits_0^{1-u} dv\frac{1}{m_c^2-2uv \kappa q-\kappa^2 u(1-u)-q^2 v(1-v)-i0}.
\end{eqnarray}
%%%%%%%%%%%%%%%%%%%%%%
%%%%%%%%%%%%%%%%%%%%%%
The amplitude (\ref{AFCNC}) corresponds to the generic amplitude describing 3BS correction to the form factor
(\ref{A0}) with the replacement of the usual quark propagator by an ``effective'' propagator describing the
charm-quark triangle $\Gamma_{cc}(\kappa,q)$
\begin{eqnarray}
\frac{1}{\mu^2-(\kappa-q)^2-i0}\to \Gamma_{cc}(\kappa,q).  
\end{eqnarray}
Important is that $\Gamma_{cc}(\kappa,q)$, similar to the usual propagator, is a quadratic function of
its momentum variables. Therefore the consideration presented in the text for the amplitude (\ref{A0})
can be directly applied to the amplitude (\ref{AFCNC}).

\end{document}